\documentclass[traditabstract]{aa}
\usepackage{graphicx} 
\usepackage[usenames,dvipsnames]{color}

\begin{document}
\title{On the narrow emission line components of the LMC novae \\ 2004 (YY Dor) and 2009a}
   \author{Elena~Mason\inst{1} and
           Ulisse~Munari\inst{2}
             }             

   \offprints{mason@oats.inaf.it}

  \institute{INAF Osservatorio Astronomico di Trieste,
              Via G. B. Tiepolo 11, 34134, Trieste\\
              \email{mason@oats.inaf.it}
      \and INAF Osservatorio Astronomico di Padova, 36012 Asiago (VI), Italy}

   \date{Received YYY ZZ, XXXX; accepted YYY ZZ, XXXX}

     \abstract{
We present early decline spectra of the two Large Magellanic Cloud novae: LMC 2004 (YY Dor) and LMC 2009a and
discuss their spectral an line profile evolution with special emphasis on the existence and appearance of a sharp 
component. We show that the narrow component that characterizes the emission lines in the maximum spectra of 
nova LMC 2004 originates in the ejecta. The He\,{\sc ii} 4686 \AA\, narrow emission which appears at the onset of
the nebular phase in both novae is somewhat controversial. Our observations suggest that the corresponding line forming
region is physically separated from the rest of the ejecta (the broad line region) and environmentally different.
However, the lack of late time observations covering the super-soft source (SSS) phase, the post-SSS phase and the
quiescence state does not allow to securely establish any non-ejecta origin/contribution as, instead, in the case of
U\,Sco and KT\,Eri.

               }

   \authorrunning{E. Mason and U. Munari}
   \titlerunning{Narrow lines in two LMC novae}

   \maketitle

\section{Introduction}

The spectra of an increasing number of classical novae (CNe) show emission
line profiles characterized by the superposition of a narrow component
($\overline{FWHM}\sim$800 km/s\footnote{The narrow line width decrease with time.}) over a ``rectangular-like'' broader
component
(FWHM$>$3000 km/s).  This narrow component usually appears in spectra
obtained during the decline from maximum, and it is detectable (when the
wavelength coverage allows it) in the H-Balmer lines and He\,{\sc ii} 4686.  The
recent discovery about the presence of such a narrow component has been made
possible by the systematic, high S/N and high resolution spectroscopic
monitoring - well into advanced decline - carried out on most of the novae
appeared during the last decade (e.g.  SMART/ATLAS program by Walter et al. 
2012, ANS Collaboration program by Munari et al.  2012, ARAS database by Buil \& Teyssier 2012). 
Its velocity, profile, intensity and timing of appearance, can be equally explained by an equatorial ring companion
to bipolar ejecta (Munari et al.  2011), an emission from the
underlying binary system (Seckiguchi et al. 1988, 1989, 1990; Walter \& Battisti 2011; Mason et al.  2012), 
or also by a projection effect of bi-conical ejecta (e.g. 
Shore et al.  2013a, Ribeiro et al.  2013a). In at least one nova (Nova Eri
2009 = KT Eri), it has been demonstrated how the He\,{\sc ii} narrow component
originated from the inner regions of the ejecta at the onset of the X-ray
super-soft phase (hereafter SSS phase for short), and from the central
binary at  somewhat later phase (Munari et al.  2014).  There are still very few
novae known to have displayed such narrow emission components, and adding a
few more to the sample is a valuable effort toward a more
comprehensive understanding of the phenomenon.

In this paper we present the results of our spectroscopic monitoring of two
novae occurred in the LMC, YY\,Dor (nova LMC 2004) and nova LMC 2009a, which
have not been discussed elsewhere and that developed a striking example of
narrow emission components.  The very high spectral resolution
(R$\geq$40000) of the data presented here allowed a clear distinction of the
narrow- and broad-components and   although we
cannot be conclusive on the ultimate evolution of the former, these novae rise interesting questions and
address future
observations and observing strategies.

   \begin{figure}
   \centering
   \includegraphics[width=8.5cm]{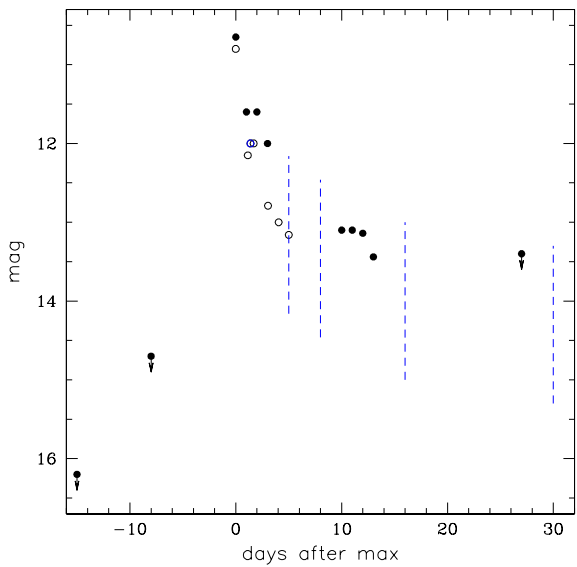}
   \caption{Light curves of YY Dor: black solid circles are for the B band
   1937 light curve (McKibben 1941; downward arrows indicate upper limits); black circles are for the V band 2004
   light curve (Liller 2004, Pearce 2004, Monard 2004, Bond et al.  2004). 
   Blue dashed vertical lines indicate the epochs of our FEROS
   observations.}
   \label{yyLC}
   \end{figure}

YY\,Dor was discovered in outburst on 2004 Oct 20.193 UT by Liller (2004)
and confirmed to be a CN by Bond et al.  (2004) and Mason et al.  (2004). 
Bond et al.  (2004) also showed that the nova matched in position nova LMC
1937 (McKibben 1941) and, therefore, that it is a recurrent nova: the second
recurrent nova observed in the LMC.  They also showed that the nova
progenitor has magnitudes B=18.8 and R=17.8, which at the distance of LMC
correspond to absolute magnitudes \footnote{with no reddening correction
applied.} M$_B$=0.32 and M$_R$=$-$0.68.  In Fig.\ref{yyLC} we compare Liller
photometric observations for the 2004 outburst with those by McKibben
obtained during the 1937 outburst.  Unfortunately both eruptions have a poor
photometric coverage, from which we estimate characteristic decline times
$t_2\simeq$4 days and $t_3>$10 days (Buscombe \& de Vaucouleurs, 1955,
list $t_3\sim$20 days), which indicate YY\,Dor as one of the fastest
known novae (e.g.  Warner 1995).

\begin{figure}
   \centering
   \includegraphics[width=8.5cm]{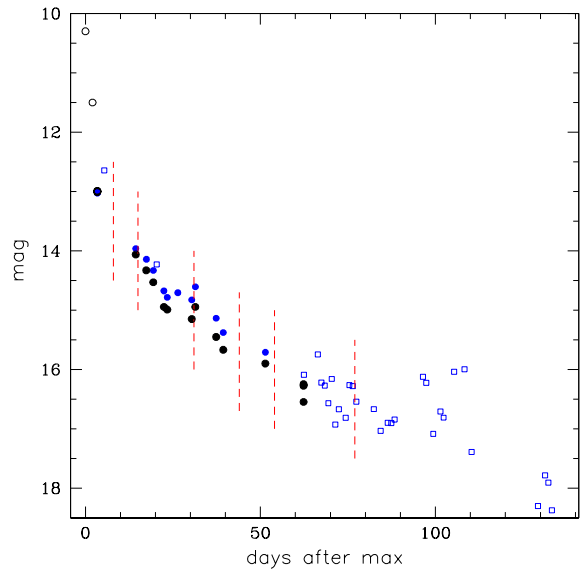}
    \caption{2009 light curve of nova LMC 2009a. Different symbols and colors
   indicate different passbands: black circles are Liller's (2009) unfiltered
   Tech.  Pan photographic discovery magnitudes; black solid circles are V
   band observations; blue solid circles are B band observations; while the
   empty blue squares are visual observations.  The visual data points
   correspond to the average of several since we have verified that
   simultaneous observations could differ by up to 1 magnitude.  With the
   exception of the two observations by Liller, all the data points are
   courtesy of AAVSO (http://www.aavso.org).  Red vertical dashed lines
   indicate the epochs of our UVES observations.}
   \label{lmcLC}
   \end{figure}

Liller (2009) discovered also the outburst of Nova LMC 2009a at unfiltered
magnitude 10.6 on Feb 5.07, 2009 UT.  He noted the proximity of this
transient to Nova LMC 1971b and suggested that the two novae could be 
two distinct eruptions of the same object.  The position of Nova LMC
1971b as given by  Graham (1971) was quite imprecise and has been revised by Shara (2000) who, re-examining the
original plates, places the nova about 2 arcmin away from nova LMC 2009a. It seems, therefore, that the two eruptions
are unrelated.  The 2009 light curve as in the AAVSO database
for the 2009 outburst is plotted in Fig.\ref{lmcLC}.  No light curve for the
1971 outburst exists with the exception of a single data point taken at the
time of the spectroscopic observation (Aug 17 1971) when it appeared at
unfiltered $\sim$13\,mag (Graham 1971).  From the AAVSO light curve we estimate
$t_2<$10 and $t_3<$20 days for this nova. Early spectroscopy of nova LMC 2009a was taken by Orio et al. 2009 and Bond
et al. 2009. 

\section{Observations and data reduction}

The log of observations of the two programs is
presented in Table\,\ref{obslog}.  Both used high resolving power
(R$\geq$40000) and spectral coverage nearly matching the whole optical range.

\begin{table}
\scriptsize
\caption{Log of observations.}             
\label{obslog}      
\centering          
\begin{tabular}{@{}l@{}c@{}c@{~~~}c@{~~~}c@{~~}c@{}}
\hline
&&&\\
nova & ``age'' & obs date & instr. & $\lambda$-range  & exptime \\ 
&  (days past max)  &  (UT) &&  (\AA) & (sec) \\
&&&\\
YY\,Dor    & 5  & 25/Oct/2004 & 2.2+FEROS & 3700-9200 & 2$\times$400  \\  
YY\,Dor    & 8  & 28/Oct/2004 & 2.2+FEROS & 3700-9200 & 5$\times$800  \\
YY\,Dor    & 16 & 05/Nov/2004 & 2.2+FEROS & 3700-9200 & 3$\times$1800 \\
YY\,Dor    & 30 & 19/Nov/2004 & 2.2+FEROS & 3700-9200 & 3$\times$2400 \\
YY\,Dor    & 67 & 26/Dec/2004 & 2.2+FEROS & 3700-9200 & 4$\times$3900 \\
&&&\\
LMC\,2009a & 8  & 13/Feb/2009 & UT2+UVES & 3200-10200 & 400           \\              
LMC\,2009a & 15 & 20/Feb/2009 & UT2+UVES & 3200-10200 & 530           \\              
LMC\,2009a & 21 & 26/Feb/2009 & UT2+UVES & 3200-10200 & 700           \\              
LMC\,2009a & 44 & 21/Mar/2009 & UT2+UVES & 3200-10200 & 2$\times$1000 \\              
LMC\,2009a & 54 & 31/Mar/2009 & UT2+UVES & 3200-10200 & 1900          \\              
LMC\,2009a & 77 & 23/Apr/2009 & UT2+UVES & 3200-10200 & 2$\times$3600 \\              
&&&\\
\hline    
\end{tabular}
\end{table}

\subsection{YY Dor}

YY\,Dor was observed on five different occasions, between 5 and 67 days past
discovery, with the ESO (now MPIA) 2.2m + FEROS.  FEROS is a fiber-fed,
cross dispersed high resolution (R=48000) echelle spectrograph mounted on a
thermally controlled bench away from the telescope (Kaufer et al.  1999). 
This guarantees that the instrument is very stable since there are no
flexure, nor variations induced by changes in the ambient temperature.  The
cross-dispersion is obtained with a prism and this eliminates second-order
contamination at red wavelengths.  FEROS delivers, within one frame, spectra
covering the wavelength range 3700-9200\,\AA, with just two small gaps
between 8534-8541 and 8861-8875\,\AA \, due to non overlapping orders. 
FEROS limitation is in the flux calibration since fiber-fed high resolution
spectrographs are usually designed for high radial velocities accuracies
only.  As the entrance aperture is circular (FEROS fiber diameter is 1.8
arcsec on the sky), the recorded signal at a given wavelength depends not
only on the seeing conditions, but also on the atmospheric dispersion which
differently displaces different wavelengths with respect to the fiber center
(i.e.  the star image will be centered or misaligned by a different factor
at different wavelengths).  This effect strongly depends upon zenith distance. 
In addition, since the telescope guides on a nearby star (which is at some
distance from the science target) and this is done at the effective
wavelength provided by the guiding camera, also differential atmospheric
refraction comes into play (see, for detailed treatment of both effects,
Donnelly et al.  1989).  Note that at the time of our YY\,Dor observations
FEROS was not yet equipped with an ADC (atmospheric dispersion corrector), so
neither the atmospheric dispersion, nor the differential refraction were
compensated for, with large flux losses at both ends of the recorded
wavelength range.  Hence, we did not attempt to flux calibrate the YY\,Dor
data.

\begin{figure*}
   \centering
   \includegraphics[height=18cm, angle=270]{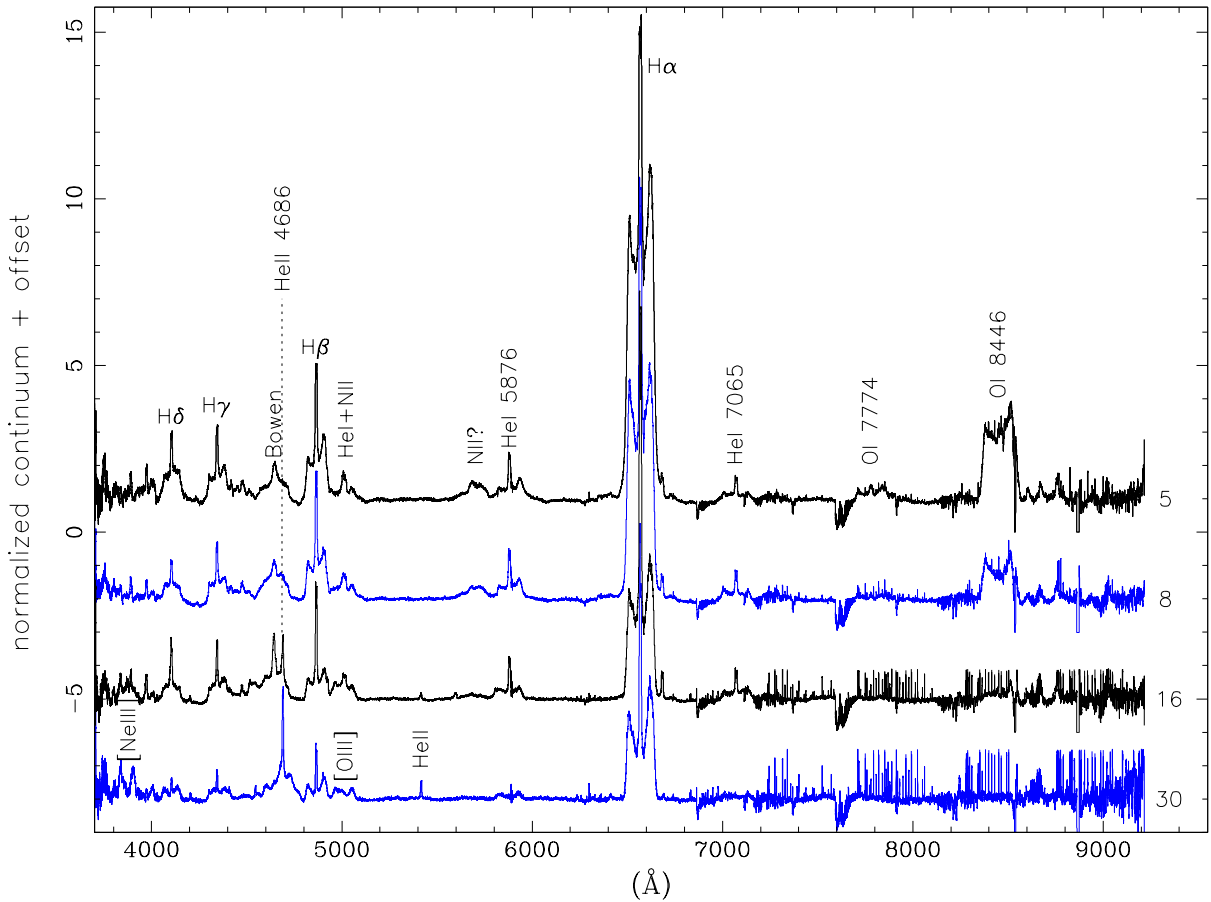}
   \caption{The sequence of our FEROS spectra showing the evolution of nova
   YY Dor 2004 outburst.  Time increase from top to bottom and the numbers
   on the right side of each spectrum indicates the spectrum age in days
   after maximum/discovery.  Spectra have been smoothed with a running
   boxcar of 33 and shifted vertically for clarity: spectrum 8 has been
   offset by -3, spectrum 16 by -6 and spectrum 30 by -9.  Note that the
   spikes on the red side of each spectrum are due to poorly subtracted sky
   emission and/or telluric absorptions.  This is particularly evident in
   relative long exposures and low SNR spectra.}
   \label{yySPCs}
   \end{figure*}

The FEROS spectra were reduced using the ESO-MIDAS instrument pipeline that is
publicly available and that applies standard pre-processing (dark and bias
subtraction in addition to flat fielding), wavelength calibration and
spectral extraction (we adopted the optimal extraction variant).  We
did not deconvolved the spectra for the instrumental resolution since all
the spectral features we detect on the nova are fully resolved: unresolved
interstellar features have widths between 4 and 7 pixels (from the blue to
the red end of the spectrum) corresponding to FWHM $<$10 km/s at any
wavelength.  The sequence of our YY\,Dor FEROS spectra is shown in
Fig.\ref{yySPCs}.  The day +67 spectrum is not included in this plot since
it is characterized by extremely low S/N.

\subsection{LMC 2009a}

Nova LMC 2009a was observed with UVES at the VLT between day +8 and +77
after discovery.  UVES is a two arms cross-dispersed echelle spectrograph
mounted at the Nasmith focus of UT2 (Dekker et al.  2000, see also the
instrument user manual available on-line).  UVES is capable of observing
with both arms simultaneously thanks to dichroic beam splitters within the
instrument optical path.  The available dichroichs are DIC1 and DIC2 which
have cross-over wavelength at 4500\,\AA\, and 5500\,\AA, respectively. 
The wavelength range covered by each arm varies with the choice of dichroic, grating and grating angle.  In order to
cover the whole optical range it is necessary to take two separate exposures with
optimal combination of dichroic, grating and central wavelength.  This is
part of the standard instrument setups offered by ESO and allows to cover
the wavelength range 3200-10200\,\AA\, with only two gaps at
5745-5840\,\AA\, and 8515-8660\,\AA\, due to the red-arm's CCD mosaic. 
UVES is not fiber-fed and its entrance aperture is a slit (one per
arm) created by two movable blades, placed after the dichroic beam splitter. 
We adopted slit widths of 1'' which delivered a resolution 
R$\simeq$40000.

Since UVES is equipped with a derotator in front of the instrument, it is
possible to compensate for the atmospheric dispersion by having it tracking
the parallactic angle during the exposures, with no need to insert an ADC
into the optical path.  This allows relative flux calibration with no color
dependent light losses.  UVES also has a slit viewer which allows to
compensate for differential atmospheric refraction effects during the
exposure.  It should be noted, however, that the VLT/UVES calibration
plan observes the spectrophotometric standard star only at the beginning or at
the end of the night, limiting the accuracy of the final flux calibration.

\begin{figure*}
   \centering
   \includegraphics[height=18cm, angle=270]{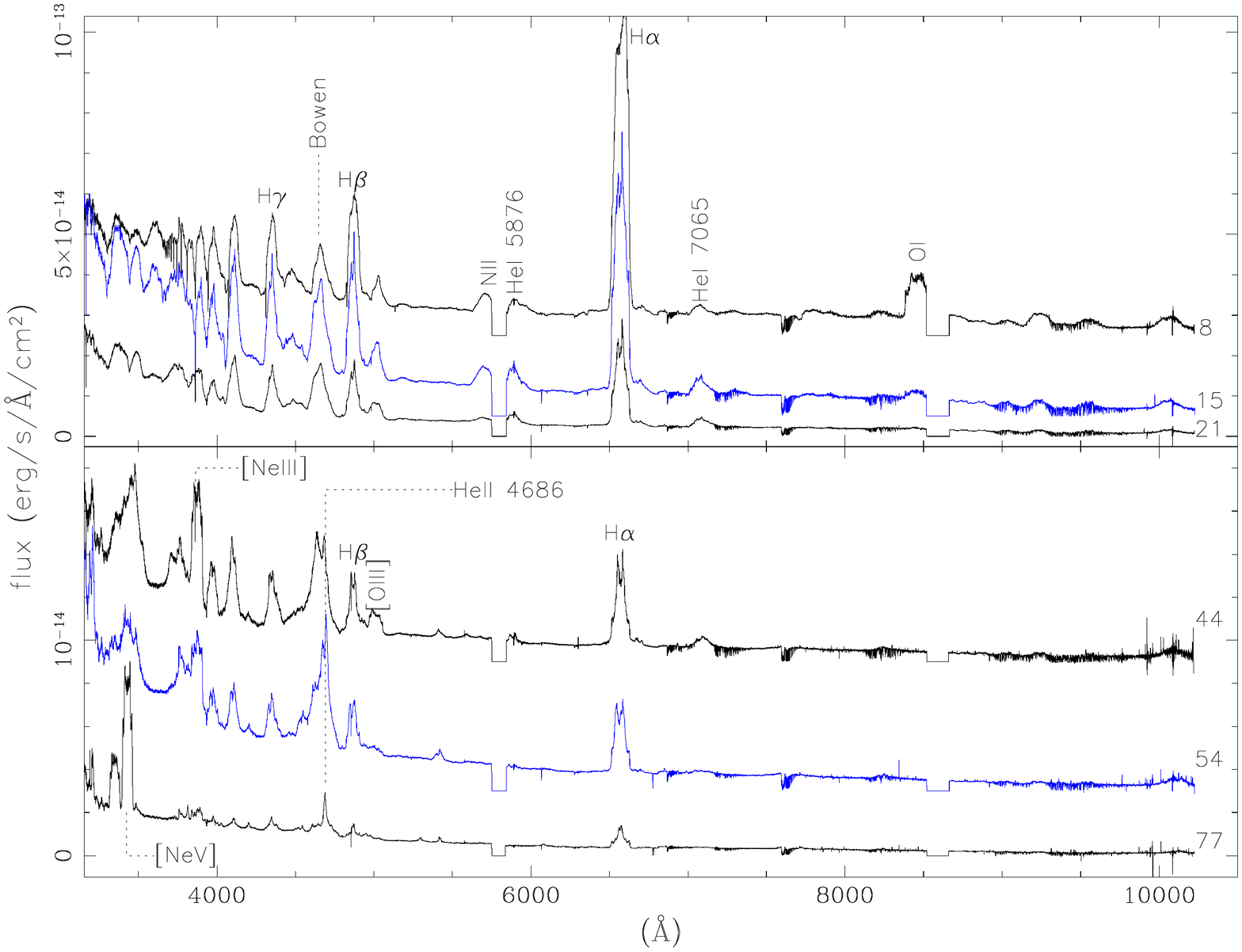}
   \caption{The sequence of UVES spectra taken for LMC 2009a. Time increase
   from top to bottom and the numbers on the right side of each spectrum
   indicates the spectrum age in days after maximum/discovery.  Spectra have
   been smoothed with a running boxcar of 33 and shifted vertically for
   clarity.  In the top panel spectrum 8 has been offset by 2.5e-14
   erg/s/cm$^2$/\AA \, and spectrum 15 by 2.5e-14 erg/s/cm$^2$/\AA.  In the
   bottom panel spectrum 44 has been offset by 0.9e-14 erg/s/cm$^2$/\AA\, and
   spectrum 54 by 0.3e-14 erg/s/cm$^2$/\AA.  The spectra are flux calibrated
   and corrected for interstellar galactic reddening as explained in the
   text.}
   \label{lmcSPCs}
   \end{figure*}

The UVES data were reduced using the instrument pipeline, available on line. 
This is based on {\sc esorex} and CPL libraries and perform image
pre-processing optimal extraction and flux calibration.  The sequence of our
UVES spectra of Nova LMC 2009a is presented in Fig.\ref{lmcSPCs}.  As for
the FEROS spectra of YY Dor, also UVES spectra of Nova LMC 2009a were not
deconvolved for the instrumental resolution, all spectral features of
interest in this paper being fully resolved.  The spectra in
Fig.\ref{lmcSPCs} have been corrected for E(B-V)=0.05 reddening
(adopting the standard $R_V$=3.1 extinction law).  This E(B-V) has been
derived from two different methods: 1) the Munari \& Zwitter (1997) relation which
correlates the Na\,{\sc i}\,D $\lambda$5890 equivalent width (EW) with the
interstellar reddening and 2) the H column density in the direction of the
LMC as mapped by the GASS and LAB projects (McClure-Griffiths et al.  2009,
Kalberla et al.  2010, Kalberla et al.  2005, Bajaja et al.  2005).  In
particular, we measure EW=0.147 from the Na\,{\sc i} absorption,
corresponding to E(B-V)=0.045 (from Munari \& Zwitter's Eqs.  2 and 3), and
a H column density of 4.36e20 cm$^{-2}$ from the H21 surveys, corresponding
to E(B-V)=0.053.  We have excluded from this latter measurement a second H21
emission at the radial velocity of the LMC since we did not observe any Na\,{\sc i}\,D absorption at that
wavelength, in our spectra. It might be that nova LMC 2009a is sufficiently offset with respect to the LAB and GASS
pointing direction to miss this second intervening cloud.  The velocity
range of the galactic Na\,{\sc i}\,D absorption and H21 emission line
brightness are in good match.

\section{The spectral evolution}

While this paper ultimately focuses on the line profile, nonetheless we
feel appropriate to briefly overview the general spectroscopic evolution
of these two novae, since it has not been presented elsewhere.

\subsection{YY Dor} 

Our monitoring of nova YY\,Dor starts when the object has already faded by
$\sim$2.5-3 mag from the observed maximum (day +5 and +8) and continues
until the early nebular phase which commenced around day +30.  The two early
decline spectra are virtually identical with only minor variations
in the relative intensities of the observed lines.  The spectra are
characterized by just a few broad emissions (FWHM$\leq$7000 km/s, FWZI$\simeq$9000-10000 km/s), among which we identify
Balmer
H lines (H$\alpha$ to H8), H-Paschen (12 to 14), O\,{\sc i} 8446 and 7774,
the Bowen 4640 blend (with N\,{\sc iii} stronger than C\,{\sc iii}),
He\,{\sc i} 7065, 6678, 5875, 5015, 4471, and N\,{\sc ii} 5001 and 5005 \AA\, (multiplet 19), 5679 \AA\, (multiplet 3). 
Si\,{\sc ii} 6347 and 6471 could be present too, while some weak lines blending with He\,{\sc i} 4471 remain 
un-identified.  We exclude Mg\,{\sc ii} since
we do not detect any significant emission in the Mg\,{\sc ii} lines $\lambda$8317
and 8328, while their identification with  O\,{\sc ii} or N\,{\sc ii} is also dubious.  The O\,{\sc i}
8446/7774 line ratio on day +5 and +8 spectra is large enough to exclude a
dominant recombination origin from O\,{\sc ii}.  Hence, fluorescence is the driving
mechanism forming O\,{\sc i} 8446 and since we do not observe O\,{\sc i} lines at 7425 and
7002\,\AA, it is not continuum fluorescence but Ly$\beta$ pumping
fluorescence.  The relative intensities H$\alpha$/O\,{\sc i}, which is much less
than $\sim$10$^3$, indicates that H$\alpha$ is optically thick and the n=2
level is over-populated (Strittmatter 1977).  While, the relative intensity
of the O\,{\sc i} 8446/7774 indicates that the emitting gas has relatively large
densities of order $n>10^{10}$ cm$^{-3}$ (see Figure 4 in Kastner \& Bhatia 1995).

The spectral evolution of YY Dor (Fig.\ref{yySPCs}) continues with the
disappearance of the O\,{\sc i} emission lines and the first appearance of the
He\,{\sc ii} $\lambda$4686, on day +16: signatures, respectively, that the
ejecta density is decreasing and the energy of the emerging ionizing
radiation is increasing.  The day +30 spectrum shows clear, though weak, traces
of both [Ne\,{\sc iii}] and [O\,{\sc iii}] emission, marking the start of the nebular phase. 
These forbidden transitions and, in particular [Ne\,{\sc iii}],  will be somewhat stronger $\sim$1.5
months later, as visible on the SMART/ATLAS spectra of
this nova (Walter et al.  2012).  Also, by day +30, He\,{\sc ii} 4686 has surpassed
in strength the H$\beta$ line, while the simultaneous disappearance of He\,{\sc i}
indicates increasing ionization of the ejecta and density bounded conditions.

\subsection{LMC 2009a}

The early decline spectral evolution of nova LMC 2009a depicted in Fig.\ref{lmcSPCs}
is similar to that of YY\,Dor described in the previous section.  The day +8
spectrum displays strong and very broad (FWHM$\sim$4500 km/s,
FWZI$\sim$7000-7500 km/s) H emission lines from both the Balmer (H$\alpha$
to H$\zeta$) and the Paschen (7 to 12) series, together with a strong O\,{\sc i}
8446 (also 7774 is visible) and the Bowen 4640 blend.  Similarly to YY Dor,
the O\,{\sc i} emissions are due to fluorescence from Ly$\beta$ pumping as it can be
concluded by applying {\bf a} similar reasoning.  The Bowen blend instead is
probably due to continuum fluorescence since we do not observe O\,{\sc iii} 3444 and
3425 (Selvelli et al.  2007, Williams \& Ferguson 1983).  He\,{\sc i} and N\,{\sc ii} are
also present.  We identify in particular He\,{\sc i} 7065, 6678, 5875, 5015,
4922, 4471 and possibly, on the bluer side of the spectrum, He\,{\sc i} 3616, 3587,
3572, 3487, 3471 and 3354, as well as N\,{\sc ii} at 5001\,\AA \, and
 possibly 5679\,\AA.  In addition, we detect Na\,{\sc i}D and the 
Fe\,{\sc ii}\ multiplet 42 lines, as narrow absorption features (FWHM $\sim$230 km/s) that are
blue-shifted by $\sim$2300 km/s with respect to the LMC rest wavelength.  Narrow
absorption features of identical characteristics are visible also in the
Balmer lines (H$\beta$ and higher terms) and possibly in O\,{\sc i} 7774 (for the latter the
uncertainty arise from the merging at this wavelength of two adjacent echelle
orders and the contamination by the nearby atmospheric telluric band).  On
day +15 the narrow absorptions have disappeared, the He\,{\sc i} emission
strengthened together with the Bowen blend, while the O\,{\sc i} weakened.  The day
+21 spectrum is apparently similar to the previous two but in fact O\,{\sc i} has
definitively disappeared and [Ne\,{\sc iii}] and [O\,{\sc iii}] first appeared, marking the
beginning of the nebular phase.  [Ne\,{\sc iii}] and [O\,{\sc iii}] reach maximum intensity on
day +44, when also He\,{\sc ii} 4686 emission appears.  He\,{\sc ii} 4686 will surpass
H$\beta$ in intensity on our day +54 spectrum.  The appearance of He\,{\sc ii}
emission occurs simultaneously with the onset of the SSS phase
in the Swift satellite observations (Schwarz et al.  2011).  Day +77 is when
the nebular transition of [Ne\,{\sc iii}] and [O\,{\sc iii}] disappear and [Ne\,{\sc v}] becomes the
strongest emission of the entire spectrum, while the He\,{\sc ii} 4686 remains the
strongest among the permitted transitions. On the day +77 spectrum we also observe emissions from O\,{\sc vi} 3811
3834 (blending with an unidentifed transition) and 5292, confirming the high ionization conditions prevailing
in the ejecta. The relative ratio of the observed O\,{\sc vi} emissions and the unconvincing presence of other
O\,{\sc vi} lines, suggest that the doublet 3811, 3834 \AA\, and the 5292\,\AA\, are fluorescence pumped by
the O\,{\sc vi} resonance lines at 150\,\AA\, and 95\,\AA, respectively, with which they share the upper level. 
The continuum has flattened indicating that it is now peaking at shorter wavelengths probably beyond
the UV, into the X-ray range.

\section{The line profiles}

There are growing evidences that nova ejecta are not spherical, but
axis-symmetric, and not uniform but clumpy, with probably a small (and
nonetheless largely uncertain) filling factor.  

The excellent FEROS and UVES spectral resolutions allow constraining both the large
and the small scale structure of the ejecta, identifying possible different ejecta component and/or characterizing
possible different physical conditions (e.g. density and opacity) within the ejecta substructures. 
Here we do not make any assumption on the geometry and the velocity field of the ejecta, but evince ejecta components
and substructures by simply analyzing the line profiles in the velocity space.

\subsection{YY Dor}

   \begin{figure*}
   \centering
   \includegraphics[width=6cm,angle=0]{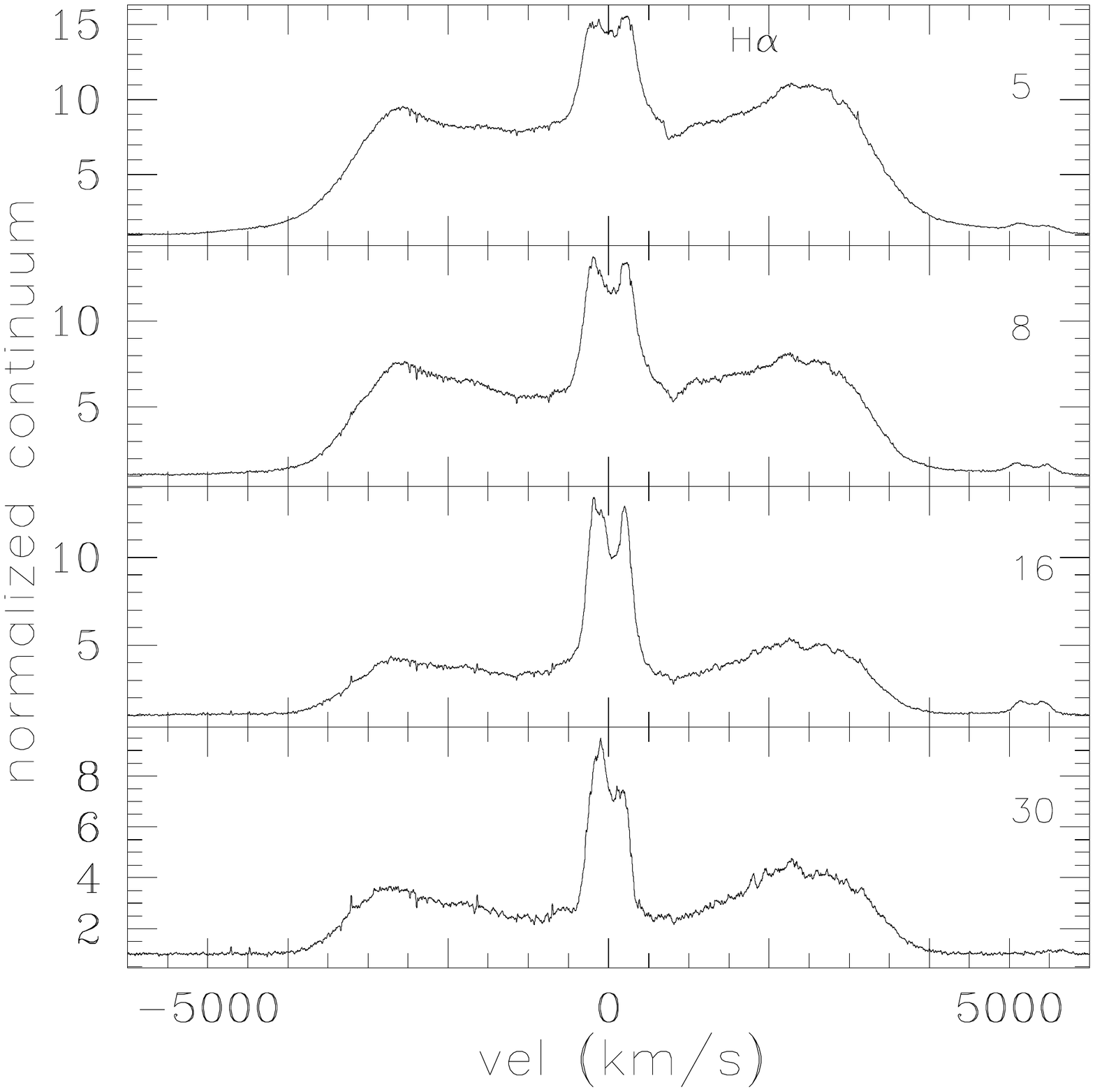}
   \includegraphics[width=6cm,angle=0]{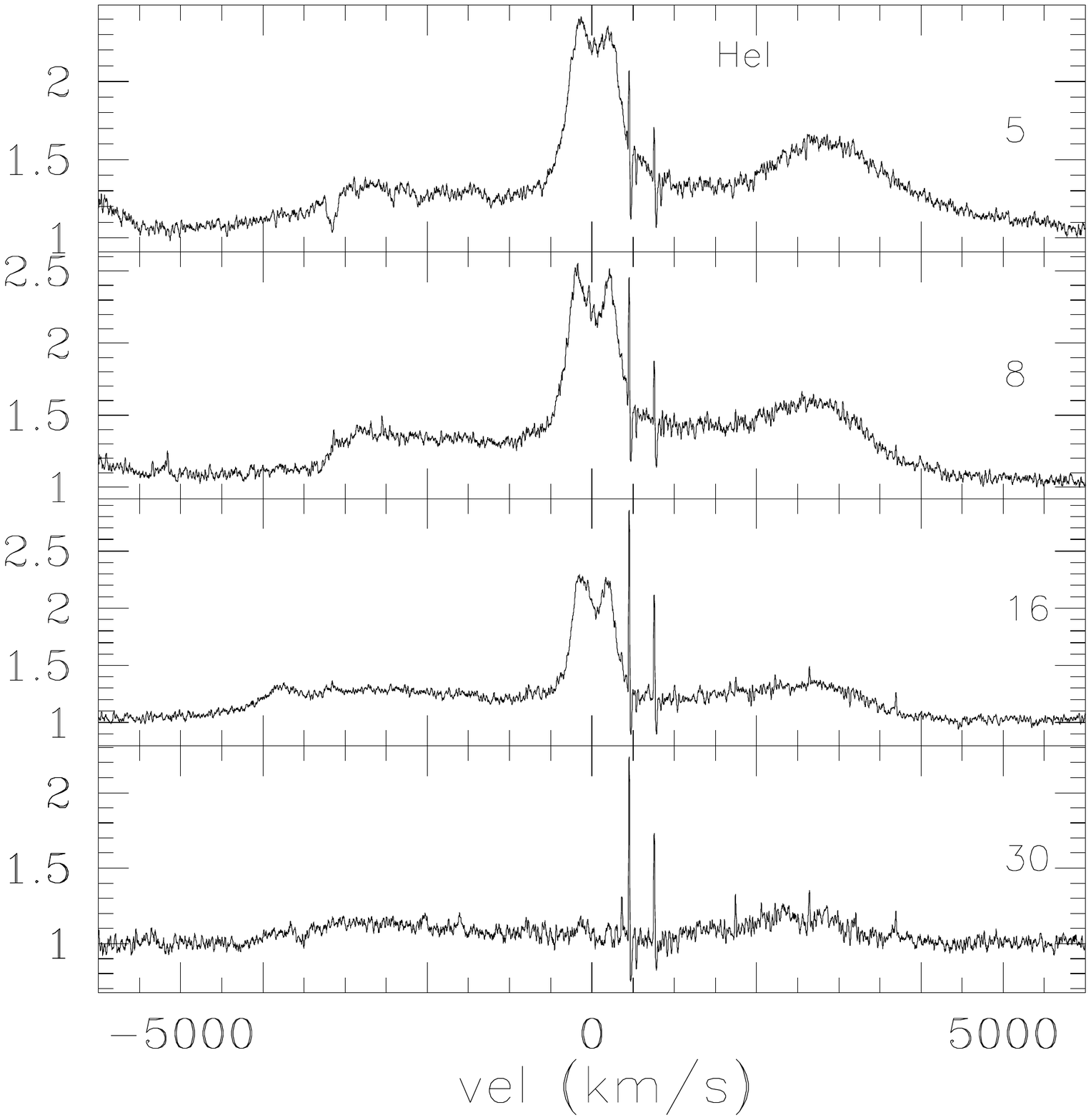}
   \includegraphics[width=6cm,angle=0]{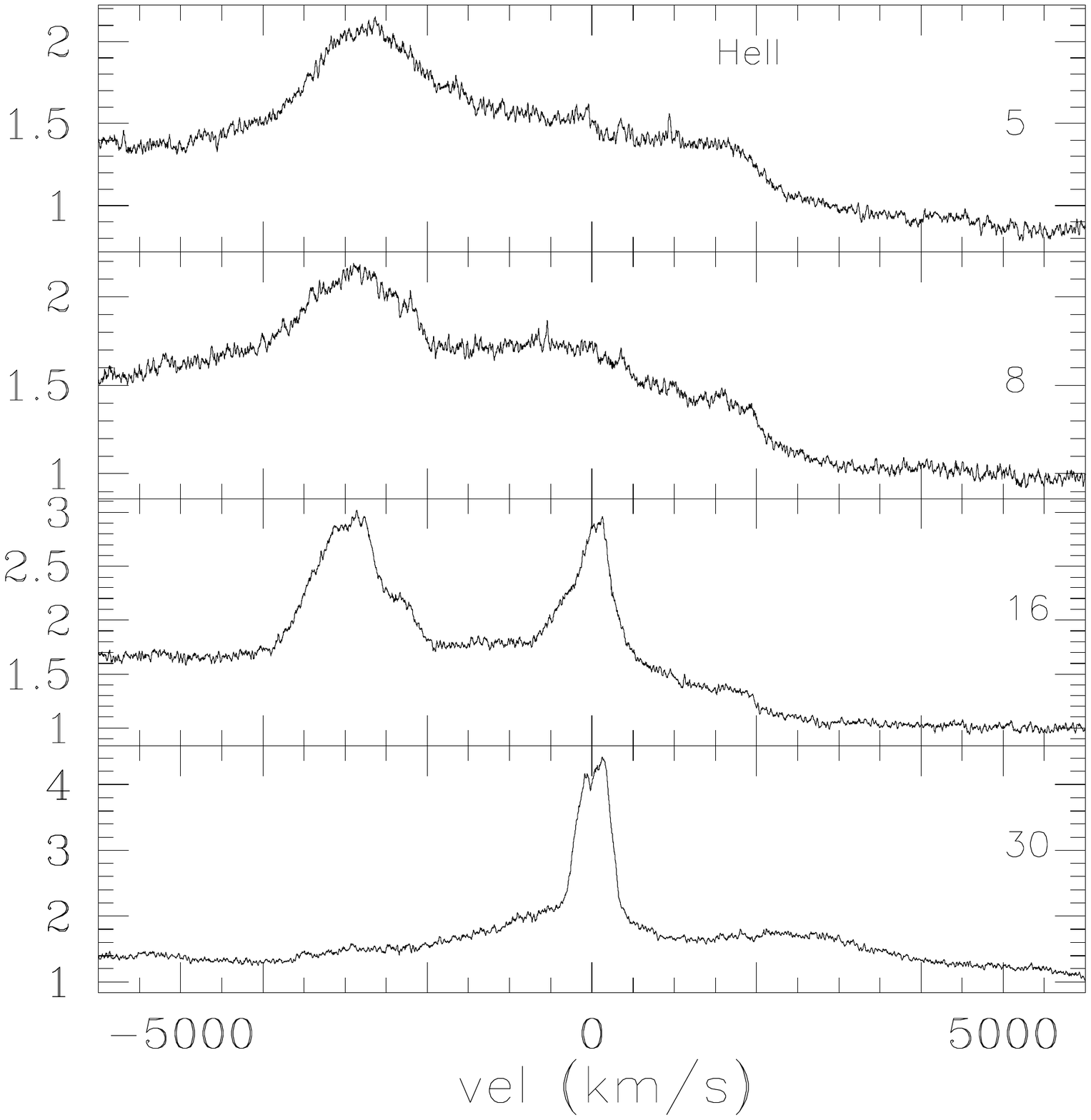}
   \caption{The profile evolution of YY Dor's H, He\,{\sc i} and He\,{\sc ii} emission lines.
   Each line identification is marked on the top of each panel.  The numbers
   on the right side of each subpanel indicate the age of the nova in day
   after maximum.}
   \label{yyEV}
   \end{figure*}

   \begin{figure}
   \centering
   \includegraphics[width=7cm,angle=0]{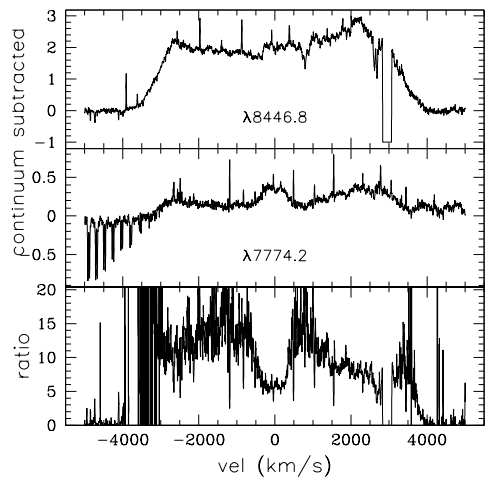}
   \includegraphics[width=7cm,angle=0]{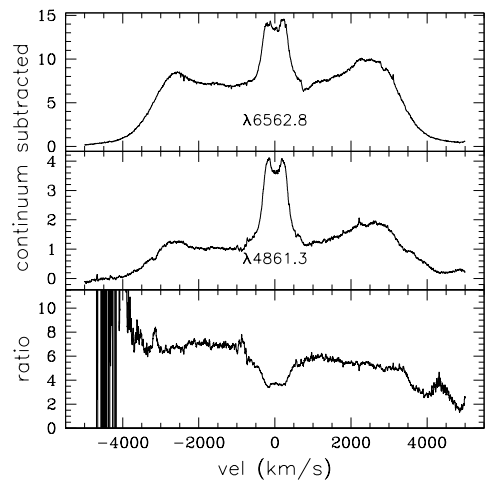}
   \caption{Top panels: YY Dor O\,{\sc i} $\lambda$8446 \& $\lambda$7774 line profiles
   and their ratio; Bottom panels: same as above but for the Balmer lines
   H$\alpha$ and H$\beta$.  See Section 4.1 for discussion.}
   \label{yyLineR}
   \end{figure}

YY Dor line profiles are peculiar since from the very maximum (see for
example the SMART/ATLAS spectrum of Oct 21 2004) they present
a double-peaked but otherwise sharp component (FWHM=700 km/s) on-top a much
broader (FWZI=9000 km/s) rectangular pedestal (e.g.  Fig.\ref{yyEV}). 
We note that these line profiles cannot result from a simple projection effect of an optically thin, single
component, axis-symmetric ejecta. First because the emission lines are optically thick at maximum (e.g. H$\alpha$, see
Sec 3.1).
Secondly because the line ratios suggest different physical conditions in the broad pedestal and the narrow emission
line. For
example the O\,{\sc i}8446/OI7774 line ratio and the H$\alpha$/H$\beta$ line ratio
(Fig.\ref{yyLineR}), show smaller values in the narrow component suggesting
larger densities (see, e.g.  Kastner \& Bhatia 2005,
Osterbrock \& Ferland 2006). In addition, the early
forbidden transitions in our day +30 spectrum consist of just the broad component, indicating that the
densities in the narrow emission component are still too large to produce
nebular transitions.

We tend to exclude an accretion disk origin of the narrow component for two basic
reasons: (a) the H-Balmer narrow emission is already present in the earliest spectra when the ejecta
are optically thick and the underlying binary is veiled by the ejecta itself, and (b) by
scaling the day +5 spectrum to the observed Rc magnitude (12.61; from SMART
database), the flux radiated by the H$\alpha$ narrow component alone is almost 
40$\times$ the flux radiated in the whole Rc band by the entire quiescent YY Dor (Rc=19.32).  This is hardly
compatible, for example, with an
enhanced accretion disk emission as it has been suggested in the past.  Nor the Balmer decrement of the narrow
components seems compatible with typical accretion disks since we measure H$\alpha$/H$\beta\sim$2.3 and
H$\gamma$/H$\beta\sim$0.55 in the scaled spectrum (to be compared with the Balmer decrements in Williams 1980). In
addition, was the double peaked profile of the narrow component forming in
an accretion disk, it would indicate a relatively high orbital inclination
and imply relatively large Doppler shifts due to the expected orbital motion. 
However, we did not detect any Doppler shift in our 4 average spectra across
30 days,  nor we found it between the consecutive exposures taken within
each observing run/epoch.

The narrow component in the He\,{\sc ii} line profile appears later and shows a
different profile with respect to the corresponding feature seen in the
other emission lines.  Initially (day +16), the He\,{\sc ii} 4686 displays a single
peaked profile and an extended blue wing, contrary to the H-Balmer and the
He\,{\sc i} narrow component which has double peaked profile and no significant
wings.  Later, on day  +30, when the He\,{\sc i} has disappeared, the narrow
component of both the Balmer and the He\,{\sc ii} lines appears single peaked but
the He\,{\sc ii} 4686 emission still shows an extended blue wing (though somewhat flatter due to the large
contrast with the narrow emission line, Fig.\ref{yyEV}), which is missing in the Balmer lines.   At later phases,
the narrow component of He\,{\sc ii} 4686 is much stronger than the corresponding one
in H$\beta$. We do not detect a He\,{\sc ii} broad component which is either null or very weak. 

The later appearance of the He\,{\sc ii} narrow emission does not necessarily imply a different
origin from the ejecta since it could be explained by a change in the ejecta physical conditions that produces
the emission lines spectrum.  Per se, the appearance of the He\,{\sc ii} is simply an
indication that the emerging photoionizing radiation is more energetic (possibly
because the underlying H-burning WD is becoming visible. Note however that X-ray observations for YY Dor do not
exist) and the ejecta become more transparent, eventually turning optically thin.
The combination of 1) the large contrast He\,{\sc ii} narrow/broad component, 2) the null ratio He\,{\sc i}
narrow/broad component and 3) the not quite nebular conditions of our day 30 spectrum, with 4) the
consideration that there is just one ionizing source (the underlying WD), suggests that the high energy photon flux is
capable of completely ionize the narrow component region but is too weak (diluted) by the time it reaches the broad
component region. I.e. the narrow component region is density bounded in He\,{\sc ii}, while the broad component region
is ionization bounded, with the two line forming regions being physically distinct and detached.   
This also implies that the narrow component line forming region is closer to the WD than the broad one and  that it is
not appearing as a low velocity emission for just a projection effect. Lower expansion velocities in the innermost
part of the ejecta are consistent with a Hubble flow type velocity law within the ejecta. 

At the same time the large He\,{\sc ii}4686/H$\beta$ does not match nebular conditions (N$_e\leq$10$^6$\,cm$^{-3}$)
and solar abundances (N(He)/N(H)$\sim$0.10-0.15) since these produce line ratios rigorously $<$1 for a range of
temperatures (e.g. Osterbrock \& Ferland 2006). Instead it suggests fairly large densities.  Gaskell \& Rojas-Lobos
(2014) could reproduce the He\,{\sc ii}4686/H$\alpha\sim$3.7 observed in a tidal disruption event (TDE), only by
adopting a
large density cloud, N$_e\sim$10$^{11}$ cm$^{-3}$, that is truncated before or at the He$^{++}$ Stromgren radius, a
geometrical configuration substantially matching the one we just described. Similarly, detailed statistical
equilibrium computations for H and He  (Bhatia \& Underhill 1986), show that large He/H line ratios, as observed in
Wolf-Rayet and O stars, result from high density and hot plasma regions (N$_e\sim$10$^{9}$-10$^{10}$ cm$^{-3}$ and
T$\sim$10$^5$ K). 
Whether our observed large He\,{\sc ii}4686/H$\beta$ is arising from a peculiar ejecta clump, suggesting
an highly inhomogeneous ejecta (in density and/or abundances), or from a dense and hot plasma region of non-ejecta
origin (e.g. gas in the primary Roche lobe, a binary common envelope, the irradiate chromosphere of the secondary star,
shock heated gas similarly to magnetic cataclysmic variables), cannot be established with the data set in hands and any
conclusion remains purely speculative. We can certainly notice that relatively strong He\,{\sc ii} emissions (He\,{\sc
ii}4686/H$\beta\simeq$1) might be observed in some novae at late phase (e.g. T\,Pyx, LMC 2012, nova Pup 2004 and NR
TrA 2008, Walter et al. 2012) suggesting that relatively large densities and/or He abundances must occurs in nova
ejecta. Still, twice as large line ratios from just a narrow component (i.e. a line substructure), as we observe, remain
intriguing and suggestive of a non-ejecta contribution. 

\subsection{LMC 2009a}

   \begin{figure}
   \centering
   \includegraphics[width=8.5cm]{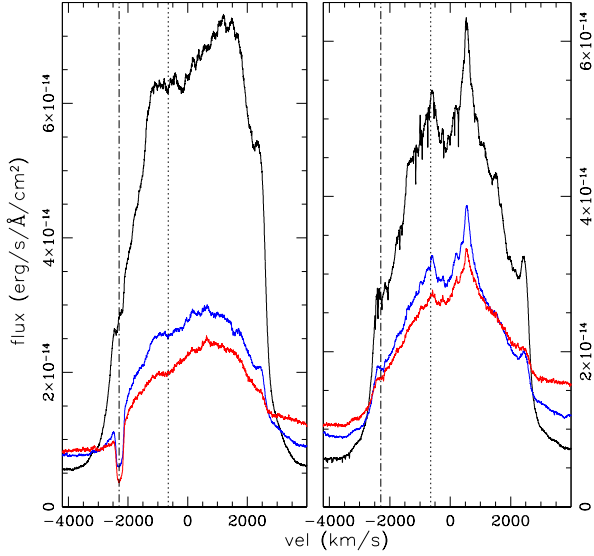}
   \caption{H-Balmer line profiles of nova LMC 2009b at day +8 (left panel)
   and +15 (right panel).  The black color is for the H$\alpha$ profile,
   blue is for the H$\beta$ line and red for H$\gamma$.  The dotted vertical
   line mark the "depression" at -650 km/s discussed in the text, while the
   dotted-dashed line marks the "pseudo P-Cyg absorption" (at -2300 km/s),
   also discussed in the text.  See section 4.2.}
   \label{maxProf}
   \end{figure}

\begin{figure}
   \centering
   \includegraphics[width=8.5cm,angle=0]{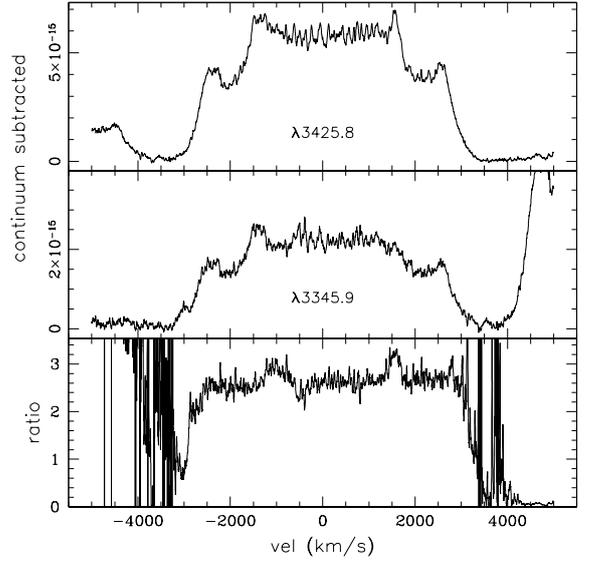}
   \caption{The Ne\,{\sc v} line profile and their ratio in day +77 spectrum showing
   the local difference across the line profile.  See Sec.4.2 for
   discussion.}
   \label{neEhe2}
   \end{figure}

The early decline spectra of LCM 2009a show broad emission lines
which results from the ``blending'' of individual substructures.  This is
evident when considering the narrow absorption and the emission profile
evolution from day +8 to +15.  In facts, the sharp and narrow ($\sim$ 230 km/s) absorption that is visible at
$-$2300 km/s in the Balmer and low ionization metals has little to null emission component in the Na\,{\sc i}\,D
and Fe\,{\sc ii} transitions and is not the blue-most feature in the Balmer lines. This can be explained either by an
extremely thin and elongated expanding shell, which however is not responsible for the bulk of the emission observed in
the H lines, or, more simply, by a clump of the ejecta that is placed between us and the nova photosphere.  
Similarly we might notice that in the spectrum of day +8 the H emission lines peak at $\sim$
1000-1500 km/s and show a depression at about -650 km/s (Fig.\ref{maxProf}). 
This latter depression turns into a relatively strong emission peak on day
+15 spectrum, suggesting a clumpy ejecta in which the clump-substructures
evolve rapidly and independently/differentially to each other
(Fig.\ref{maxProf}).  Also the nebular transitions and the [Ne\,{\sc v}] doublet in the
day +77 spectrum support this scenario since their flux ratio does not show a
flat profile in velocity space but the clear presence of substructures
(Fig.\ref{neEhe2}).

\begin{figure}
   \centering
   \includegraphics[width=8.5cm,angle=0]{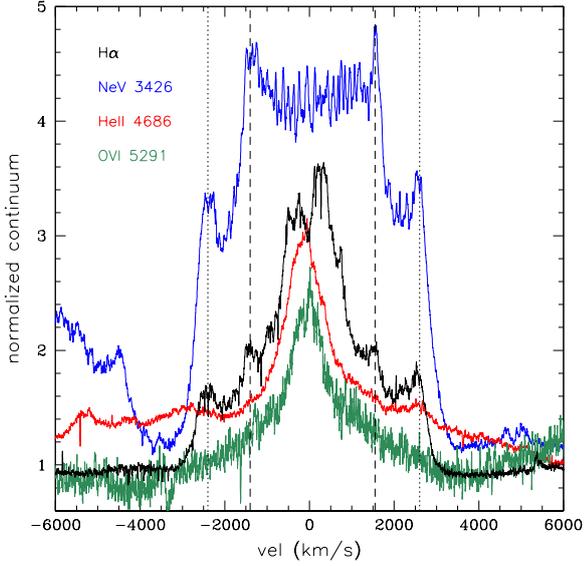}
   \includegraphics[width=8.5cm,angle=0]{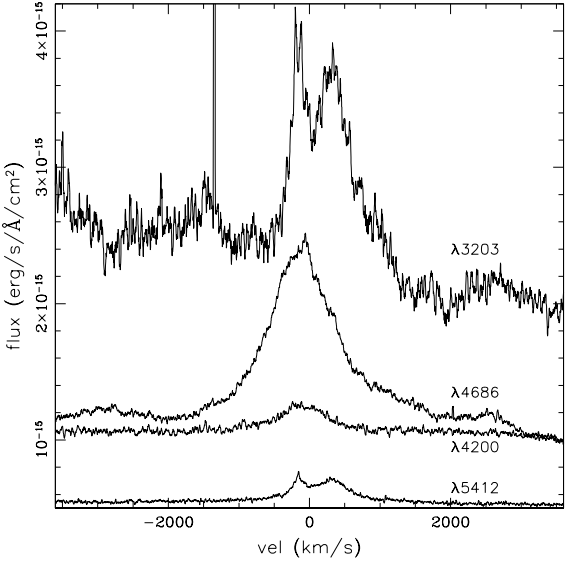}
   \caption{Top Panel: The different line profiles of four different
   transitions from four different ions as observed in the day +77 spectrum. 
   Vertical dot and dashed lines marks the width of the two broad
   components.  Bottom Panel: The He\,{\sc ii} line profiles in day +77 spectrum. 
   See text (Section 4.2) for more details.}
   \label{lprof}
   \end{figure}

At the same time, not all emission lines share the same large scale profile. 
We have said before that the emission lines are very broad but in fact not
all are and, as the nova evolves into the nebular phase, a multicomponent line
profile emerges.  The nebular emission lines in the +44/+77 day spectra and
the O\,{\sc i} 8446 in the day +8 spectrum show substantially identical profiles
consisting of two superposed rectangular components of apparent width
$\sim$5000 and 3000 km/s, respectively.  This two components profile is
interesting since it well matches the projection effect of a bipolar ejecta
(cones or dumbbell) which have been recently identified and/or modeled in a
number of novae (e.g. T\,Pyx, Shore et al. 2013b; Mon 2012, Shore et al.  2013a, Ribeiro et al.  2013b;
Cyg 1992, Vel 1999, Shore et al.  2013a; KT Eri, Ribeiro et al.  2013a; LMC 2012, Schwarz private communication) 
 supporting the idea of an axis-symmetric bi-conical polar ejecta
common to all or most CNe. The H-Balmer lines, too, in the day +77 spectrum
show such a double component profile (Fig.\ref{lprof}), however, they also show a
third narrower component on top of the two rectangular emissions. On the other side, we see only a narrow emission with
broad wings 
 in He\,{\sc ii} and O\,{\sc vi} lines (Fig.\ref{lprof}).

Few things shall be noted about such a nebular spectrum. First, in the epochs preceding out last observation, i.e.
between day 44 and 54, the lines portion within $\pm$2000 km/s from the center shows dramatic profile variations. Day
44 roughly matches the start of the SSS phase (Schwarz et al. 2011) and this, as for many other novae (Ness 2012), 
is characterized by strong variations before reaching a stable X-ray emission. In a clumpy ejecta, different
clumps might be seen as weaker or stronger in a given transition and at a
given time depending on their position with respect to the observer, their
density and opacity conditions and their distance from the radiation source. 
They might also appear weaker or stronger at different given times depending
on the radiation travel time, and their expansion/dilution time scale in the circumbinary space. This can explain
our observed line profile variations at the onset of the X-ray emission. Similarly, the narrow He\,{\sc ii} and O\,{\sc
vi} emissions could match a dense (to emit only permitted transitions) 
clump which is intercepting a sufficient number of high energy photons.
However, again, the large He(4686)/H$\beta$ flux ratio ($\geq$3.98) suggests fairly high densities, likely of the order
10$^{9}$-10$^{11}$\,cm$^{-3}$ (e.g. Ganskell \& Rojas-Lobos 2014, Bhatia \& Underhill 1986); while, the broad wings are
possibly inconsistent with an ejecta clump. We shall also notice that the He\,{\sc ii} transitions display
different line profiles (Fig.\ref{lprof}). These might be explained with contributions from multiple regions,
each with different physical conditions, or with opacities and optical depth effects though these are very much
unexpected in typical astrophysical contexts and in particular within a passive ejecta.

We conclude noting that, independently on its actual nature and origin, certainly the narrow line forming
region has physical conditions which are dramatically different from that of the broad line either because of density,
or abundance or both. In facts, both regions display high ionization potential energy ions (Ne\,{\sc v} IP energy is
$\sim$126 eV, to be compared with that of O\,{\sc vi} and He\,{\sc ii}: $\sim$ 138 and 54 eV, respectively), however,
the broad line forming region show forbidden transitions and the narrow does not. While, applying a nebular analysis to
the day 44 and 77 spectra in the attempt to estimate the He/H abundances in the broad and the narrow line forming
region, respectively, we obtain values that differ by a factor of at least 2.

\section{The narrow component in the literature}

As mentioned in the introduction section, the number of objects for which
there is the claim of an observed narrow component is increasing.  To put
some order and to the aim of better isolating consistently similar phenomena
we define as {\em novae showing narrow emission line components} those for
which (1) the narrow component sits on-top of a clearly distinct and much
broader pedestal, (2) it is seen in He\,{\sc ii} 4686, and (3) with time the He\,{\sc ii}
4686 narrow component overtakes, in strength, that displayed by H$\beta$. 
To our knowledge, the novae fulfilling such a criterion are: the two LMC
presented here, U Sco (Barlow et al.  1981, Sekiguchi et al.  1988, Munari
et al.  1999, Diaz et al.  2010, Mason et al.  2012, Anupama et al.  2013),
V394 Cra (Sekiguchi et al.  1989, Williams et al.  1991), LMC 1990b
(Sekiguchi 1990, Williams et al.1991), KT Eri (Munari et al.  2014), DE Cir
(Walter et al.  2012).  V2672 Oph, which has a line profile very similar to
YY Dor, cannot be included in the list since it has been observed only in
H$\alpha$ (Munari et al.  2011, Walter et al.  2012).  Among these novae,
the appearance of the narrow He\,{\sc ii} emission matched the start of the SSS
phase (Schwarz et al.  2011) in LMC 2009a, U\,Sco, and KT\,Eri.  Therefore,
on one side it is tempting to suggest that the appearance of the He\,{\sc ii} 4686
is the optical marker of the start of the SSS phase in all the objects
listed above.  On the other side, while the simple observation of the narrow
He\,{\sc ii} emission cannot provide information upon the location of the line
forming region, it stays true that the observations of the SSS phase
indicates that the binary is no longer veiled by the ejecta and it is
virtually observable.  It is also inevitable to note that in both U\,Sco and
KT\,Eri at advanced decline, the He\,{\sc ii} emission has been shown to originate from the binary and
not from the ejecta (Thoroughgood et al.  2001, Mason et al. 
2012, Munari et al.  2014).  We might note that the hypothesis of a binary
origin and in particular of an accretion disk origin for the He\,{\sc ii} narrow
component in U\,Sco, V394 CrA, and nova LMC 1990b was already put forward by
Skiguchi et al.  (1988, 1989 and 1990) who suggested an enhanced accretion
disk emission and, following Webbink et al.  (1987), interpreted the
emission as due to a face-on accretion disk.  Sekiguchi et al. were aware of
the problems posed by their model, confirmed by the later discovery that 
U\,Sco is an eclipsing system. We also know, however, that U\,Sco He\,{\sc ii} narrow emission originates
in the (already reformed) accretion disk (Thoroughgood et al.  2001). 
Unfortunately only U\,Sco has been observed in time resolved spectroscopy up
to now; and only U\,Sco and KT\,Eri have to been followed years after
outburst (Hanes 1985, Munari et al.  2014) showing that the narrow emission
and the He\,{\sc ii} persists, well after the ejecta (broad component) dissolved.

For the outburst of future novae, it appears of great relevance to follow
their evolution as long as possible, well into their nebular phases, so
to monitor the evolution of any narrow component to the aim of establishing:
\begin{enumerate}
\item whether the narrow component ever develops a nebular/forbidden
transition, proving a definitive ejecta origin;
\item  whether, and in case, when the contribution to the narrow component
from the central binary overtakes that from the ejecta (similarly to KT Eri
and U\,Sco);
\item whether or not it is signature of restored accretion (as in U\,Sco);
and
\item whether the presence and properties of the narrow component correlates
with basic nova parameters like type, decline speed, mass and velocity of
the ejecta, ...
\end{enumerate}

 We can only address these issue adopting strategic and focused
observations at the next suitable occasions.  However, for all of the
objects above, it remains critical to observe and possibly monitor them now
in quiescence, to verify whether the narrow He\,{\sc ii} emission persists as the
strongest emission and for how long.  This would provides indication on the
ionizing radiation emitting source, its cooling, if any, and possibly the
line formation mechanism.  In facts, though it is not an exclusive observing
evidence, the strongest He\,{\sc ii} emissions are typically observed in magnetic CVs
rather than in disk systems (e.g. Warner 1995).

We have mentioned in the introduction section that YY Dor is a recurrent nova.  The careful reader has certainly noticed
that of the seven novae listed above, four are recurrent novae and three are somewhat
debated (i.e.  LMC 2009a, KT Eri and DE Cir, Jurdana-Sepic et al.  2012, Pagnotta \&
Schaefer 2014).  Without entering into the discussion on the possible short
recurrent nature of these novae and the recently debated definition of
recurrent novae and classes of recurrent novae, we just note here that few
of these seven novae show additional common phenomenology such as the
expansion velocities (all FWHM are in excess of $\sim$3500 km/s), a [Ne] dominated
nebular spectrum (U Sco, Diaz et al.  2011, Mason et al.  2012; YY Dor,
Walter et al.  2012; LMC 2009a, this paper, and DE Cir, Walter et al. 
2012), slightly evolved secondaries (U Sco, V394 CrA, KT Eri and LMC 2009a,
Darnley et al.  2012; for the case of KT Eri see also Munari \& Dallaporta
2014) and orbital period of the order of 1 day (U Sco, Schaefer 2010; V394
CrA, Schaefer 2010, and LMC 2009a, Bode et al.  2009) which are worth to
explore since they suggest possible profound common physical characteristics
(e.g.  massive ONe primary).

\begin{acknowledgements}
We acknowledge with thanks the variable star observations from the AAVSO
International Database contributed by observers worldwide and used in this
research.  We thank Robert Williams for his careful reading of the manuscript and his valuable suggestions and
comments. We also thanks the anonymous referee for the provocative requests which allowed to strengthen our
paper. EM is deeply grateful to Pierluigi Selvelli and Steven Shore for their help, the stimulating discussions and
the confrontations. 
\end{acknowledgements}

\end{document}